# Diachronic Linked Data: Towards Long-Term Preservation of Structured Interrelated Information

Yannis Stavrakas
IMIS / RC "Athena"
yannis@imis.athena-innovation.gr

George Papastefanatos
IMIS / RC "Athena"
gpapas@imis.athena-innovation.gr

Theodore Dalamagas
IMIS / RC "Athena"
dalamag@imis.athena-innovation.gr

Vassilis Christophides
Computer Science Department, University of Crete
christop@csd.uoc.gr

## ABSTRACT
The Linked Data Paradigm is one of the most promising technologies for publishing, sharing, and connecting data on the Web, and offers a new way for data integration and interoperability. However, the proliferation of distributed, interconnected sources of information and services on the Web poses significant new challenges for managing consistently a huge number of large datasets and their interdependencies. In this paper we focus on the key problem of preserving evolving structured interlinked data. We argue that a number of issues that hinder applications and users are related to the temporal aspect that is intrinsic in linked data. We present a number of real use cases to motivate our approach, we discuss the problems that occur, and propose a direction for a solution.

## 1. INTRODUCTION
More and more corporate, government, or even crowd-sourced data are published on the so-called *Web of Data* and become available for potential data consumers outside their production site. Open Data[1] published according to the *Linked Data*[2] *Paradigm* [2] are expected to play a catalyst role in the way structured information is exploited in the large scale, and offers a great potential for building innovative applications that create *new value* from the already collected data. Although it may originate from heterogeneous sources, Linked Open Data (LOD) may be published in the RDF data model along with their meaning defined in various ontologies, and thus can be freely interlinked, forming in such a way *a global space of shared data* [3]. In this respect, LOD technology is expected to provide a substrate for the discovery of new knowledge that is not contained in any individual source, and the solution of problems that were not originally anticipated by their creators.

However, little attention has been given to the fact that the LOD Cloud is subject to *change under no central control*. Only recently the scientific community has started to study the evolution of the Semantic Web [8] and the dynamics of Linked Data [5]. Analogous studies have been made in the past in the context of the Web of Documents [9] and Curated Databases [10]. The bulk of the research efforts conducted so far has been focused on scalable RDF data stores and efficient SPARQL query engines in centralized and distributed settings (see [11] for a recent survey) as well as on automated methods for ontology matching and alignment [12]. We believe that innovative technologies are needed for supporting the full lifecycle of evolving LOD on the Web of Data: from data extraction, transformation and integration, to change monitoring, quality assessment and repair, until synchronization and long-term preservation. Such technologies are expected to foster sustainable LOD ecosystems by improving decision making, ensuring transparency in data processing, adopting common policies to privacy-aware data sharing, curation and preservation policies, while minimizing rework.

The rest of the paper is organised as follows: In section 2, we introduce the problems that, in our view, need to be addressed. In section 3, we describe a set of motivating use cases that require support of diachronic LOD, and in section 4 we discuss the particularities and the challenges for preserving evolving LOD data. In section 5 we present the main aspects of the proposed framework, and we conclude the paper in section 6.

## 2. THE LOD ECOSYSTEM
Linked Data in conjunction with Web 2.0 technologies have transformed the Web from a publishing-only environment into a vibrant information place where yesterday's passive readers have become active data collectors and generators themselves. The Web of Data is essentially a social system involving several players: besides *data producers* and *consumers* third parties may additionally contribute. For instance, *data matchmakers* try to discover publicly available data silos (not always in a linked data format) for particular application needs and sometimes to reconcile the encountered discrepancies at the schema or instance levels by establishing mappings / correspondences commonly used in a domain of interest. In this context the Linked Data value chain becomes far more complex [4] than in traditional enterprise or scientific applications.

The above roles can be played by various public or private entities (persons, enterprises, associations, and research institutes). This multiplicity of roles and entities together with the overwhelming availability and dynamicity of data sources introduce new problems.

(a) How can we *monitor* changes of third-party LOD datasets released in the past (the *evolution tracking problem*) or how

---
[1] en.wikipedia.org/wiki/Open_data

[2] linkeddata.org

can newly released versions be considered by ongoing data analysis processes (the *change synchronization* problem)?

(b) How can we *understand the evolution* of LOD datasets w.r.t. the real world entities they describe (the *provenance* problem) and how various data imperfections (e.g., granularity inconsistencies) can be repaired (the *curation* problem)?

(c) How we can *assess* the quality (temporal and spatial) of harvested LOD datasets in order to be able to decide which and how many versions of them deserve to be further preserved (the *appraisal* problem)?

(d) How do we *cite* particular versions of a LOD dataset (the *citation* problem), and how will we be able to retrieve them when looking up a reference in the form in which we saw it – not the most recently available version (the *archiving* problem)?

(e) How can we *spread preservation costs* to ensure long-term access even when the original motivation for publishing has changed (the *sustainability* problem)?

Our aim is to address these crucial questions for *structured*, *interlinked* and *evolving* information such as Linked Open Data (LOD) published on the Web of Data or intranets. Our approach is centered on the notion of *diachronic linked data* for which *time* become intrinsic. Diachronic linked data is, by its nature, aims to be *self-preserving*: it records its own history, its evolution and its production and usage context. The aim of this paper is to describe the core elements of a preservation platform which enhanced by a number of tools facilitate *data governance* among different data producers, consumers and matchmakers. According to this vision both the data and metadata are diachronic, and thus the need for third-party linked data preservation (e.g. by memory institutions) will be greatly reduced.

## 3. MOTIVATING USE CASES

In this section we briefly describe a number of real use cases that demonstrate how different aspects of the same core problems are encountered in a variety of applications and data domains.

**Scientific data**. Biology research communities produce, consume, and archive rapidly large amounts of data. Scientific communities like that rely increasingly on the Web for collaboration, through the publication and integration of experimental and research results. Moreover, scientists in those communities would often like to review how and why the recorded data have evolved, in order to compare and re-evaluate previous and current conclusions. Such an activity may require a search that moves backwards and forwards in time, spreads across various databanks, and performs complex queries on the semantics of the changes that modified the data. In those cases, simply revising past document snapshots and differences between versions may not be enough.

As a concrete example, consider a major provider of *in vitro* diagnostic laboratory equipment who has to optimize the workflow of the diagnostic chain in the laboratory. This workflow includes communication of information between successive laboratory systems, not necessarily of the same generation, each one relying on its own embedded version of the dataset of micro-organisms and related reagents (drugs and chemicals used for tests). The state-of-the-art of micro-organism taxonomy and known effects of reagents being changing quite fast, all those local datasets have different reference versions of knowledge. To synchronize efficiently the information workflow and prevent errors of interpretation in a critical environment where the patient's life is often engaged, the global laboratory information system has to store (a) successive versions of the various reference sources, (b) the knowledge that each component of the laboratory has of the version used, and (c) correspondence rules between previous versions and the current ones. This information control and workflow system is running in a closed proprietary environment, but public reference information such as biological taxonomy, genetic information, drug nomenclature etc. will be more and more available as linked open data. So the information system needs to be synchronized also with the evolution of those external sources.

**Authority data**. The Publications Office of the European Union is in charge of maintaining and making available for institutional customers and the general public *authority tables* about entities such as countries and their administrative subdivisions, languages, currencies, community corporate bodies and organisations, and legislative procedures. All those "authorities" are described at a very fine-grained level of attributes, including in particular various codes and types of labels (long name, official name, acronym …), in each official language of the European Union. Those entities and their attributes are likely to change over time, therefore most information has to be annotated with a valid time span stamp. The information managed in the back-office, and delivered to the various users, must therefore be marked in a way that it can be filtered by time, in order to provide at any time the current authoritative vocabulary to use in official publications, but also to enable entity extraction in documents published at a certain point in time using the names and other attributes which were valid at the time of publication.

Given the official status and mission of the publishing authority, those publications should be a reference for time-aware search and information aggregation. This will be achieved not only by providing reference entities and data, but a data model simple enough to be re-usable by organizations facing similar maintenance and publication issues.

**Governmental data**. An innovative "data matchmaker" company collects publicly available data from various sources: open data from the PSI (Public Sector Information), data from social network, data from private sources (free or not), internal data from customers, and web data. Most of this data is in tabular form, and most are associated with space or time: they represent temporal series, spatial series, or both. One major problem is the maintenance of this collected data over time. Current experience shows that from the data collected during one year, 25% disappears (the address or the structure changes, but most of the time, the data simply vanishes). For the data that changes over time, the company scans the sources at a frequency close to that of the change frequency, and then, depending on the semantics of the data, performs one of the following: either update the data set by keeping the last values only, or store the sequence of data observed over time.

The company is very interested in getting a good definition and understanding of the correctness of data over time, together with strategies and algorithms for maintaining a correct data repository over time.

## 4. LINKED DATA PECULIARITIES AND CHALLENGES

Data diachronicity has been studied in the past under different data management research topics such as version management, change detection and modeling, temporal data management, data and schema evolution, data provenance, data archiving and preservation. However, existing open preservation frameworks proposed for scientific and cultural data [1] cannot cope with the intrinsic features of linked data that introduce a number of new challenges.

- **Linked Data are Structured**: LOD *stewards*[3] and *custodians*[4] need to manage not just individual statements but entire collections of RDF triples that may additionally satisfy certain quality criteria (e.g., integrity constraints). Furthermore, such collections may be *interconnected* since individual entities described in a particular Web data source could refer through typed links to other related entities in the LOD space so forming *graph-shaped data spaces*. Exploratory browsing, querying and matching techniques are needed in order to assist LOD *matchmakers* in the arduous task of discovering, interlinking and most importantly preserving LOD collections from autonomous data sources distributed all over the Web. This calls for effective crawling, entity extraction and ranking techniques in LOD spaces as well as efficient entity co-reference methods to explicitly or implicitly refer to other relevant LOD collections to an analysis task. In particular, efficient *entity resolution algorithms* have to be devised to reduce the necessary number of comparisons between LOD collections of potential interest.

- **Linked Data are Dynamic**: LOD spaces are evolving *worldwide under no central authority* as new real word entities are considered for analysis or old ones became obsolete for further analysis or even due to corrections of erroneous conceptualizations actually employed by RDF triples. Unlike the setting where data are bounded by a closed work in which change monitoring is build-in, linked data are mostly hosted in the open space of the Web and their changes can be either partially observed periodically through crawling or explicitly be communicated via notification mechanisms [7]. As they are interconnected (through copying or referencing), changes need to be propagated from one LOD collection to another within or across consumer communities. Clearly, there is a need for tools assisting LOD *curators*[5] in understanding and managing the changes of evolving LOD. In particular, discovering LOD differences (deltas) and representing them as complex objects – first class citizens with structural, semantic, temporal and probabilistic characteristics has been proved to be vital in various tasks such as the *synchronization* of autonomously developed LOD versions, or *recording* and *visualizing* the evolution history of a particular LOD collection of triples. This calls for change detection tools able to produce *deltas* that can be interpreted both by *humans* and *machines*, and modeling methods that accommodate changes in LOD structuring, archiving and querying. In this way, curators can easily provide both LOD *accountability* (examine how and why changes took place in the past) and LOD *long-term accessibility (*ensure for future users access the most recent LOD version). Besides that, we also need support for declarative change querying and longitudinal queries that span across different versions of evolving LOD. Just as the Web of Data is a globally distributed dataspace, handling of changes should be done in a distributed fashion. There will be many different publishers and consumers (such as agents, indexer, consolidator platforms, etc.) of datasets with different requirements and capabilities. A distributed approach can cope with this challenge in a cost- and performance-efficient way. Finally, we need appropriate mechanisms for minimizing the impact of changes (e.g., on SPARQL queries) and enabling their self tuning to evolution operations.

- **Linked Data are Uncertain**: As LOD usage is generalized, their quality may be compromised by various forms of *data imperfections* (e.g., impreciseness, unreliability incompleteness) due to fundamental limitations of the underlying measurement infrastructures which produce them, the inherent ambiguity in the domain of interest, or even when privacy-preserving applications modify data by adding perturbations to it. Similarly, when LOD are produced by extracting structured information from text, or entity resolution algorithms are employed in sensor and social data, the results are approximate and uncertain at best. *Uncertainty* is a state of limited knowledge, where we do not know which of two or more alternative statements are true. In this respect, representing declaratively RDF triples uncertainly and answering queries over probabilistic RDF graphs is a challenging problem.

- **Linked Data are Distributed**: production, matchmaking and consumption of linked data are spread worldwide on the Cloud. A collection of rdf triples may explicitly or implicitly refer to other relevant collections. Therefore, we need to preserve not just than individual data collections, but a network of interlinked ones and connected. The problem of distributed preservation must be tackled, in the expectation that it will reduce the dependence on central data centers, whose longevity is subject to economic forces and which, in some cases, have proved unsustainable[6].

In our view, we need open specifications and tools for preserving and providing diachronic linked data that involve actors from the entire value chain of linked data. The preservation policies defined by producers, and the data needs specified by consumers will be taken into account by third party agents providing linked data preservation services. This "production – matchmaking – consumption - preservation" cycle will maximize the use of the information and the benefit coming out of it.

## 5. A FRAMEWORK FOR DIACHRONIC LINKED DATA

To address the needs of the previously described application scenarios and challenges, we need to develop a *distributed*, *service-based* infrastructure for *curation* and *preservation* of LOD through their entire lifecycle. Such a system will need to comprise the following essential functionality:

---

[3] en.wikipedia.org/wiki/Data_stewards

[4] en.wikipedia.org/wiki/Data_custodian

[5] en.wikipedia.org/wiki/Data_curation

[6] www.dcc.ac.uk/news/arts-and-humanities-data-service-decision

- *Adaptive focused crawling*. Gather linked data from the Web about a domain, together with relevant background information that is required to put the data into context. The crawler will take into account the "preservation policies" provided by the data producers, will make decisions on which links to follow first, and will dynamically adapt its frontier accordingly.

- *Change detection*. Identify changes by pulling out and comparing snapshots, or by monitoring the actions of the user. The description of each change together with any superimposed information about the change will be stored in the archive. Changes will trigger a notification mechanism that will identify related nodes and propagate the fact of the change to all possibly affected information objects.

- *Multiversion archiving*. Automatically archive each new "release" of the data, following a distributed approach for storing information. The archived data will be replicated in several nodes in order to increase efficiency and guarantee the availability and preservation of information.

- *Longitudinal query capabilities*. Answer questions efficiently with complex conditions on the provenance and evolution of information objects. It will be possible to express snapshot queries on previous instances of the data and their relationships, but also longitudinal queries that cut across snapshots to give insight about the *how*s and *why*s of the current state of information.

- *Provenance support.* Since in the LOD cloud RDF triples are usually replicated, to assess various forms of data quality, such as trustworthiness, reputation and reliability it is crucial to determine the origins of published LOD worldwide. This essentially calls for representing and reasoning on the provenance of LOD, as they are transformed by declarative SPARQL queries or inferred through logic programs. Instead of computing each possible annotation – such as trust scores– independently during data sharing, an alternative approach is to record abstract provenance information for capturing the relationship among source and derived data along with the query operators that were involved in the derivations. This provenance information can then be materialized in the repository when the data is imported and used later to compute annotations "on the fly", based on annotations on source data and how they were combined through query operators for a particular application.

Towards this direction, we propose a framework for diachronic linked data, called *L2D* (standing for Lifecycle of Linked Data). The framework is not intended to replace existing standards and tools, but rather to complement, integrate, and co-exist with them by building on previous efforts of the Linked Data community. We envisage the L2D cloud, which will be part of the Linked Open Data cloud[7], consisting of *L2D nodes*. Each L2D node will accommodate the *L2D Platform* that will integrate and provide a set of services to support Linked Data throughout their lifecycle.

Figure 1 depicts the overall architecture of the L2D Platform.

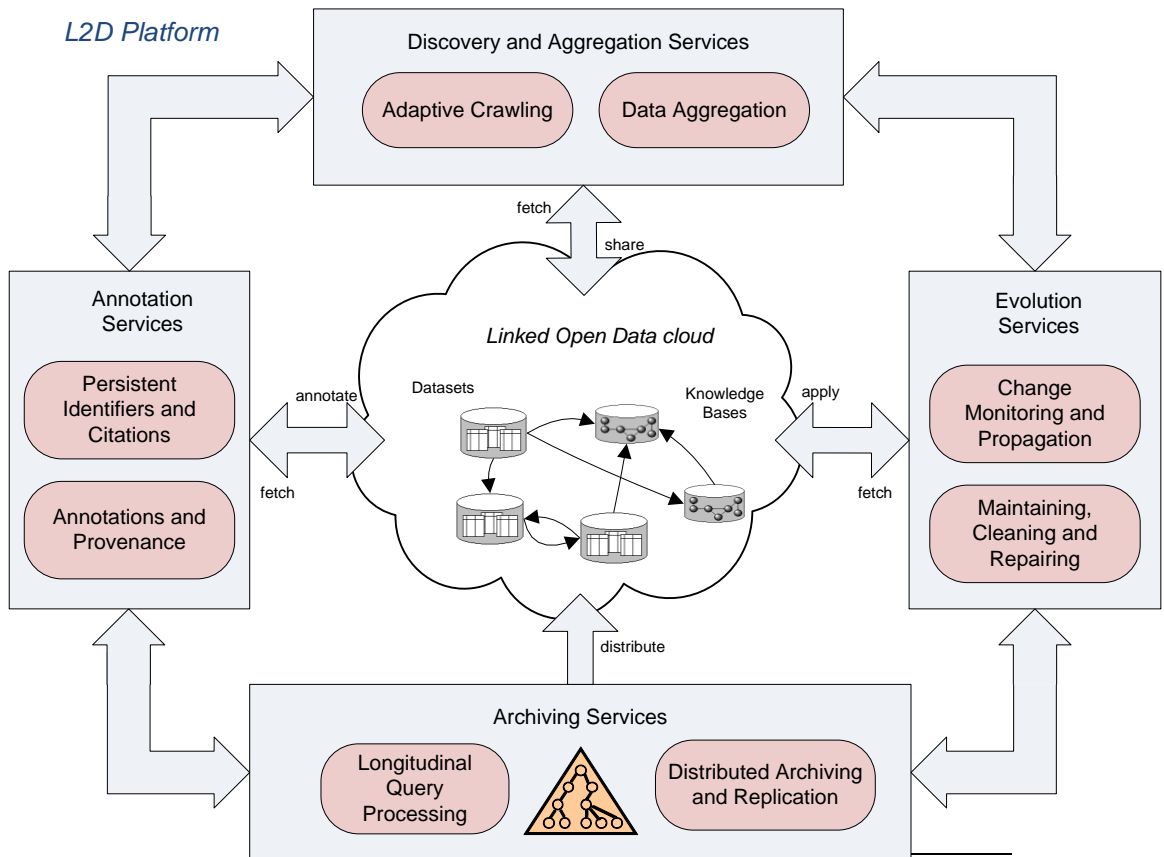

**Figure 1: Overview of the L2D Platform**

We foresee four groups of services for *long-term LOD accessibility and usability*: the *discovery and aggregation* services, the *annotation* services, the *evolution* services, and the *archiving* services, and are discussed in detail in what follows.

The *discovery and aggregation* services are responsible for detecting new relevant data from various domains, supporting the creation of new links between them, and providing input to the *annotation* and *evolution* services. They consist of the following services:

- *Adaptive crawling*. This service will monitor regularly specific domains for changes, and discover new thematically focused data, according to a specification and a set of policies set by the parties involved. A key problem is the ammount of irrelevant information often returned by focused crawlers. To address this problem, crawling should be adaptive, and decide at run time which directions to follow among the possible options in the crawler queue.
- *Data aggregation*. This service will support the creation of new links between data, by identifying information objects that correspond to the same entity. Probabilistic techniques may be used to cater for uncertainty when matching entities.

The *annotation services* are responsible for collecting superimposed information concerning the provenance, interpretation, and use of Linked Data, and then storing it to the L2D archive. They consist of the following services:

- *Persistent identifiers and citations*. This service will provide persistent identifiers for information objects. Persistent identifiers are indispensable for tracking the evolution and provenance of data, metadata, and relations between them. Moreover, the service will provide "persistent citations" [13], i.e. references to pieces of data and metadata that do not "break" in case those data are modified or removed. The lack of persistent citations is a major drawback in the Web today.
- *Annotations and provenance*. This service will record relationships between data and metadata in disparate information systems, in order to represent and store the lineage of information objects and the lineage of their relationships. It will be possible for users to contribute annotations that indicate the usage and interpretation of data. Overall, the service will provide support for understanding where a piece of information came from, and how it should be interpreted and used. Together with the service "maintaining, cleaning and repairing", which is discussed later, it will enable an explanation as to how and why a piece of data has its current form.

The *evolution services* are responsible for identifying and managing changes within the L2D cloud, and for recording those changes in the L2D archive. They consist of the following services:

- *Maintaining, cleaning and repairing*. This service will be responsible for updating the current state of a L2D node. Modifications will be applied to the operational system, while additional superimposed information will be generated, describing these modifications. The additional information produced by L2D will allow one to follow the evolution of an information object backwards or forward in time. Emphasis will be given to the treatment of changes as first class citizens [6] at the same level as information objects. Moreover, the service will offer synchronization facilities between interdependent parts of datasets and knowledge bases in the L2D cloud that will be triggered on specific events.
- *Change monitoring and propagation*. In contrast to the previous service which requires human input, this service will identify changes by pulling out and comparing snapshots of the operational system. The description of those changes will then be stored in the L2D archive. Changes may trigger a notification mechanism that will identify related L2D nodes and propagate the fact of the change to all possibly affected information objects. Probabilistic techniques will be employed where appropriate to cater for uncertainties in change discovery and in their interdependencies.

The *archiving services* are responsible for storing and accessing the data produced by the annotation services and the evolution services. They consist of the following services:

- *Archiving*. L2D will provide a service for archiving evolving Linked Data. Archiving will be automatic – with each "release" of the data – and will be efficient. L2D will follow a distributed approach for storing information. The archived data will be replicated in several nodes in order to increase efficiency, and guarantee the availability and preservation of information.
- *Longitudinal query processing*. This service will enable to efficiently answer questions with complex conditions on the provenance and evolution of information objects. It will be possible to express snapshot queries on previous instances of the data and their relationships, but also longitudinal queries that cut across snapshots to give insight about the hows and whys of the current state of information.

The *L2D Platform* will integrate these services into a cohesive framework and will be accessible not only directly from the users, but also from applications that would like to exploit the potential of individual services and components.

## 6. CONCLUSIONS

In this paper we argued that a wide range of users and applications would benefit from a framework for managing the preservation of evolving linked data ecosystems. In our view, the temporal aspects should be considered explicitly in the design of algorithms and tools for managing linked data. We presented a number of use cases from various domains to demonstrate the real need for evolution and preservation support. We discussed a number of problems we consider as closely related, and we proposed a high level architecture of a framework that would tackle those problems.